\documentclass[structabstract, longauth, letter]{aa}  

\newcommand{\ttt}[1]{\times10^{#1}}

\renewcommand{\epsilon}{\varepsilon}

\usepackage{graphicx}

\usepackage{color}
\usepackage{fixltx2e}

\usepackage{txfonts}
\usepackage{caption}
\usepackage{subcaption}

\usepackage{natbib}
\bibpunct{(}{)}{;}{a}{}{,} 

\begin{document}

\title{Discovery of TeV gamma-ray emission from the pulsar wind nebula 3C 58 by MAGIC}

%
\author{
J.~Aleksi\'c\inst{1} \and
S.~Ansoldi\inst{2} \and
L.~A.~Antonelli\inst{3} \and
P.~Antoranz\inst{4} \and
A.~Babic\inst{5} \and
P.~Bangale\inst{6} \and
J.~A.~Barrio\inst{7} \and
J.~Becerra Gonz\'alez\inst{8,}\inst{25} \and
W.~Bednarek\inst{9} \and
E.~Bernardini\inst{10}
% ...
}

\institute { IFAE, Campus UAB, E-08193 Bellaterra, Spain
\and Universit\`a di Udine, and INFN Trieste, I-33100 Udine, Italy
\and INAF National Institute for Astrophysics, I-00136 Rome, Italy
\and Universit\`a  di Siena, and INFN Pisa, I-53100 Siena, Italy
\and Croatian MAGIC Consortium, Rudjer Boskovic Institute, University of Rijeka and University of Split, HR-10000 Zagreb, Croatia
\and Max-Planck-Institut f\"ur Physik, D-80805 M\"unchen, Germany
\and Universidad Complutense, E-28040 Madrid, Spain
\and Inst. de Astrof\'isica de Canarias, E-38200 La Laguna, Tenerife, Spain
\and University of \L\'od\'z, PL-90236 Lodz, Poland
\and Deutsches Elektronen-Synchrotron (DESY), D-15738 Zeuthen, Germany
\and ETH Zurich, CH-8093 Zurich, Switzerland
\and Universit\"at W\"urzburg, D-97074 W\"urzburg, Germany
\and Centro de Investigaciones Energ\'eticas, Medioambientales y Tecnol\'ogicas, E-28040 Madrid, Spain
\and Institute of Space Sciences, E-08193 Barcelona, Spain
\and Universit\`a di Padova and INFN, I-35131 Padova, Italy
\and Technische Universit\"at Dortmund, D-44221 Dortmund, Germany
\and Unitat de F\'isica de les Radiacions, Departament de F\'isica, and CERES-IEEC, Universitat Aut\`onoma de Barcelona, E-08193 Bellaterra, Spain
\and Universitat de Barcelona, ICC, IEEC-UB, E-08028 Barcelona, Spain
\and Japanese MAGIC Consortium, Division of Physics and Astronomy, Kyoto University, Japan
\and Finnish MAGIC Consortium, Tuorla Observatory, University of Turku and Department of Physics, University of Oulu, Finland
\and Inst. for Nucl. Research and Nucl. Energy, BG-1784 Sofia, Bulgaria
\and Universit\`a di Pisa, and INFN Pisa, I-56126 Pisa, Italy
\and ICREA and Institute of Space Sciences, E-08193 Barcelona, Spain
\and Universit\`a dell'Insubria and INFN Milano Bicocca, Como, I-22100 Como, Italy
\and now at: NASA Goddard Space Flight Center, Greenbelt, MD 20771, USA and Department of Physics and Department of Astronomy, University of Maryland, College Park, MD 20742, USA
% ...
}


\date{Received ... / Accepted ... Draft version \today}

 
  \abstract
   {The pulsar wind nebula (PWN) 3C 58 is one of the historical very high-energy (VHE; $E>$100 GeV) $\gamma$-ray source candidates. It is energized by one of the highest spin-down power pulsars known (5\% of Crab pulsar) and it has been compared with the Crab nebula because of their morphological similarities. This object was previously observed by imaging atmospheric Cherenkov telescopes (Whipple, VERITAS and MAGIC), although it was not detected, with an upper limit of 2.3 \% Crab unit (C.U.) at VHE. It was detected by the {\it Fermi} Large Area Telescope (LAT) with a spectrum extending beyond 100 GeV.}
   {We aim to extend the spectrum of 3C 58 beyond the energies reported by the Fermi Collaboration and probe acceleration of particles in the PWN up to energies of a few tens of TeV.}
   {We analyzed 81 hours of 3C 58 data taken in the period between August 2013 and January 2014 with the MAGIC telescopes. }
   {We detected VHE $\gamma$-ray emission from 3C 58 with a significance of $5.7\sigma$ and an integral flux of 0.65\% C.U. above 1 TeV. According to our results, 3C 58 is the least luminous VHE $\gamma$-ray PWN ever detected at VHE and has the lowest flux at VHE to date. The differential energy spectrum between 400 GeV and 10 TeV is well described by a power-law function d$\phi$/d$E$=$f_0$($E$/1 TeV)$^{-\Gamma}$ with $f_{0} = (2.0 \pm 0.4_{\text{stat}} \pm 0.6_{\text{sys}})\ttt{-13} \text{cm}^{-2} \text{s}^{-1} \text{TeV}^{-1}$ and  $ \Gamma = 2.4 \pm 0.2_{\text{stat}}\pm 0.2_{\text{sys}}$. The skymap is compatible with an unresolved source.}
   {We report the first significant detection of PWN 3C 58 at TeV energies. We compare our results with the expectations of time-dependent models in which electrons upscatter photon fields. The best representation favors a distance to the PWN of 2 kpc and far-infrared (FIR) values similar to cosmic microwave background photon fields. If we consider an unexpectedly high FIR density, the data can also be reproduced by models assuming a 3.2 kpc distance. A low magnetic field, far from equipartition, is required to explain the VHE data. Hadronic contribution from the hosting supernova remnant (SNR) requires an unrealistic energy budget given the density of the medium, disfavoring cosmic-ray acceleration in the SNR as origin of the VHE $\gamma$-ray emission.}


   \authorrunning{Aleksi\'c et al.}
   \titlerunning{Discovery of VHE $\gamma$-ray emission from 3C 58 by MAGIC}
   \maketitle
%


\section{Introduction}
\label{intro}

The supernova remnant 3C 58 (SNR G130.7+3.1) has a flat radio spectrum and is brightest near the center, therefore it was classified as a pulsar wind nebula \citep[PWN; ][]{PWN_classification}. It is centered on PSR J0205+6449, a pulsar discovered in 2002 with the {\it Chandra} X-ray observatory \citep{Discovery_X-rays}. It is widely assumed that 3C 58 is located at a distance of 3.2 kpc \citep{Distance_old}, but a new H I measurement suggests a distance of 2 kpc \citep{Kothes2013}. The age of the system is estimated to be $\sim$ 2.5 kyr \citep{Chevalier} from the PWN evolution and energetics. However, the historical association of the PWN with the supernova of 1181 \citep{SN1181} and different measurements such as the velocity of the optical knots \citep{Optical_knots}, neutron star cooling models \citep[proposing that an exotic cooling mechanism must operate in this pulsar;][]{Slane2002}, the proper motion of the pulsar \citep{Pulsar_proper_motion}, and radio expansion of the nebula \citep{7kyr} derive ages ranging from 0.8 kyr up to 7 kyr. The pulsar has one of the highest spin-down powers known ($\dot E$ = 2.7$\times$10$^{37}$erg s$^{-1}$). The PWN has a size of 9$^{\prime}$$\times$6$^{\prime}$ in radio, infrared (IR), and X-rays \citep{Size, Size_X-rays, Slane2004, IR_Mag_field}. Its luminosity is $L_{\text{ 0.5 -- 10 keV}}=2.4\times 10^{34}$ erg s$^{-1}$ in the X-ray band, which is more than 3 orders of magnitude lower than that of the Crab nebula \citep{X_Luminosity}. 3C 58 has been compared with the Crab because the jet-torus structure is similar \citep{Slane2004}. Because of these morphological similarities with the Crab nebula and its high spin-down power (5\% of Crab), 3C 58 has historically been considered one of the PWNe most likely to emit $\gamma$ rays.

The pulsar J0205+6449 has a period  $P$=65.68 ms, a spin-down rate $\dot P=1.93\times10^{-13}$s s$^{-1}$, and a characteristic age of 5.38 kyr \citep{Discovery_X-rays}. It was discovered by the {\it Fermi}-LAT in pulsed $\gamma$ rays. The measured energy flux is $F_{\gamma>0.1 \rm \text{GeV}}$=(5.4$\pm$0.2)$\times$10$^{-11}$ erg cm$^{-2}$s$^{-1}$ with a luminosity of $L_{\gamma>0.1 \rm \text{GeV}}$=(2.4$\pm$0.1)$\times$10$^{34}$ erg s$^{-1}$, assuming a distance for the pulsar of 1.95 kpc \citep{Distance_pulsar}. The spectrum is well described by a power-law with an exponential cutoff at E$_{\text{cutoff}}$=1.6 GeV \citep{SecondPulsarCatalog}. No pulsed emission was detected at energies above 10 GeV \citep{FermiAbove10}. In the off-peak region, defined as the region between the two $\gamma$-ray pulsed peaks (off-peak phase interval $\phi$=0.64--0.99), the Fermi Collaboration reported the detection of emission from 3C 58  \citep{SecondPulsarCatalog}. The reported energy flux is (1.75$\pm$0.68)$\times$10$^{-11}$erg cm$^{-2}$s$^{-1}$ and the differential energy spectrum between 100 MeV and 316 GeV is well described by a power-law with photon index $\Gamma=1.61\pm0.21$. No hint of spatial extension was reported at those energies. The association of the high-energy unpulsed steady emission with the PWN is favored, although an hadronic origin related to the associated SNR can not be ruled out. 3C 58 was tagged as a potential TeV $\gamma$-ray source by the Fermi Collaboration \citep{FermiAbove10}.

The PWN 3C 58 was previously observed in the very high-energy (VHE; E $>$ 100 GeV) range by several imaging atmospheric Cherenkov telescopes. The Whipple telescope reported an integral flux upper limit of 1.31$\times$10$^{-11}$ cm$^{-2}$s$^{-1}$  \citep[$\sim$ 19 \% C.U. at an energy threshold of 500 GeV;][]{WhippleULs}, and VERITAS established upper limits at the level of 2.3 \% C.U. above an energy of 300 GeV \citep{3C58_Aliu_VERITAS}. MAGIC-I observed the source in 2005 and established integral upper limits above 110 GeV at the level of 7.7$\times$10$^{-12}$ cm$^{-2}$s$^{-1}$  \citep[$\sim$4 \% C.U.;][]{MAGICULs}. The improved sensitivity of the MAGIC telescopes with respect to previous observations and the {\it Fermi}-LAT results motivated us to perform deep VHE observations of the source.


 \section{Observations}

MAGIC is a stereoscopic system of two imaging atmospheric Cherenkov telescopes situated in the Canary island of La Palma, Spain (28.8$^\circ$N, 17.9$^\circ$ W at 2225 m above sea level). During 2011 and 2012 it underwent a major upgrade of the digital trigger, readout systems, and one of the cameras \citep{Upgrade}. The system achieves a sensitivity of (0.71 $\pm$ 0.02)\% of the Crab nebula flux above 250 GeV in 50 hours at low zenith angles \citep{Performance2013}.

MAGIC observed 3C 58 in the period between 4 August 2013 to 5 January 2014 for 99 hours, and after quality cuts, 81 hours of the data were used for the analysis. The source was observed at zenith angles between 36$^\circ$ and 52$^\circ$. The data were taken in \emph{wobble-mode} pointing at four different positions situated 0.4$^\circ$ away from the source to evaluate the background simultaneously with 3C 58 observations \citep{Wobble}.


\section{Analysis and results}
\label{results}

The data were analyzed using the MARS analysis framework \citep{MARS}. For every event, we calculated the size of the image (number of photoelectrons (phe) in the image of the shower) and the angular distance $\theta$ between the reconstructed arrival direction of the gamma-ray and the source position. We determined the arrival direction combining of the individual telescope information using the Disp method \citep{MARS}. A random-forest algorithm was used to estimate a global variable, the hadronness, which was used for background rejection \citep{RF}. The energy of a given event was estimated from the image sizes of the shower, its reconstructed impact parameter, and the zenith angle with Monte Carlo (MC) filled look-up tables. We optimized the cut parameters for detecting a 1\% C.U. point-like source on an independent Crab nebula data sample by maximizing the Li \& Ma significance \citep[Eq. 17, hereafter LiMa]{LiMa}. We selected events with a $\theta^2$ angle $<$ 0.01 deg$^2$, hadronness $<$ 0.18, and  size in both telescopes $>$ 300 phe. For each pointing we used five off-source regions to estimate the background. We used MC $\gamma$ rays to calculate the energy threshold of the analysis, defined here as the peak of the true energy distribution. To calculate the flux and spectral energy distribution (SED), the effective area of the telescopes was estimated using the MC simulation of $\gamma$-rays. The SED was finally unfolded using the Schmelling method to account for the energy resolution and energy bias of the instrument \citep{Unfolding}. 

\begin{figure}[!t]
 \centering
 \includegraphics[width=0.5\textwidth]{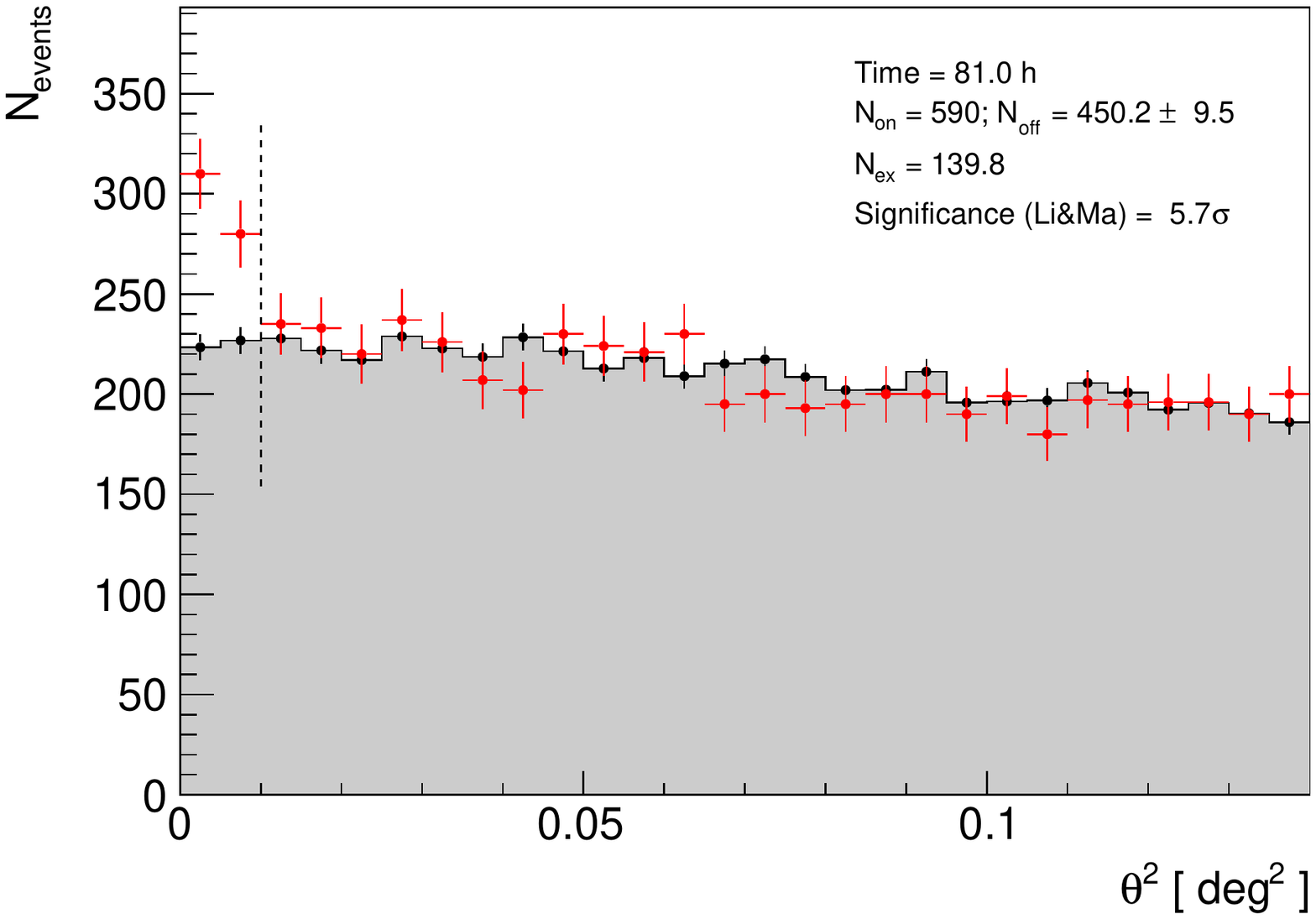}
\caption{Distribution of squared angular distance, $\theta^2$, between the reconstructed arrival directions of gamma-ray candidate events and the position of PSR 0205+6449 ({\it red points}). The distribution of $\theta^2$ for the OFF positions is also shown ({\it gray filled histogram}). The vertical dashed line defines the signal region ($\theta^2_{\text{cut}}$=0.01 deg$^2$), $N_{\text{on}}$ is the number of events in the source region, $N_{\text{off}}$ is the number of background events, estimated from the background regions and $N_{\text{ex}}$=$N_{\text{on}}$-$N_{\text{off}}$ is the number of excess events.}
  \label{Theta2}
 \end{figure}

The applied cuts and the zenith angle of the observations yield an energy threshold of 420 GeV. The significance of the signal, calculated with the LiMa formula, is $5.7\sigma$, which establishes 3C 58 as a $\gamma$-ray source. The $\theta^2$ distribution is shown in Figure \ref{Theta2}. As the five OFF positions were taken for each of the wobble positions, the OFF histograms were re-weighted depending on the time taken on each wobble position.

\begin{figure}[!t]
 \centering
 \includegraphics[width=0.5\textwidth]{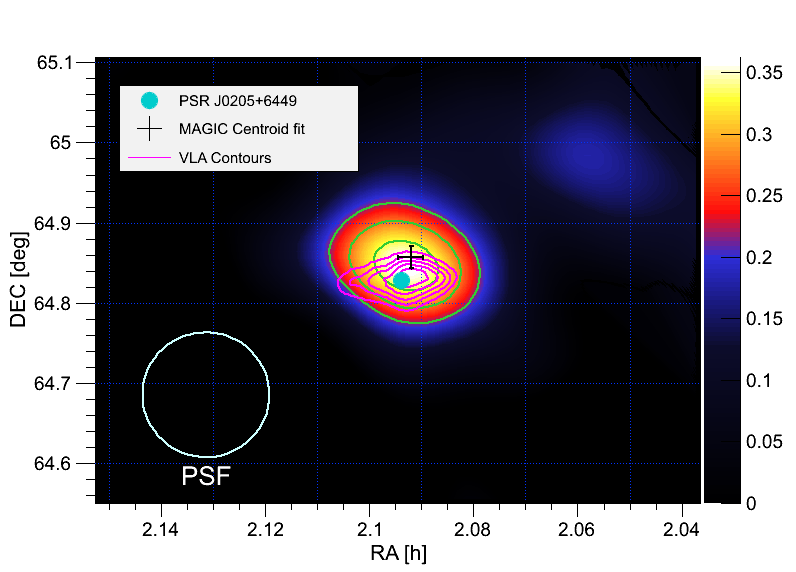}
   \caption{Relative flux (excess/background) map for MAGIC observations. The cyan circle indicates the position of PSR J0205+6449 and the black cross shows the fitted centroid of the MAGIC image with its statistical uncertainty. In green we plot the contour levels for the TS starting at 4 and increasing in steps of 1. The magenta contours represent the VLA flux at 1.4 GHz \citep{VLA}, starting at 0.25 Jy and increasing in steps of 0.25 Jy.}\label{Skymaps} 
  \label{SED}
 \end{figure}

We show in Figure \ref{Skymaps} the relative flux (excess/background) skymap, produced using the same cuts as for the $\theta^2$ calculation. The test statistics (TS) significance, which is the LiMa significance applied on a smoothed and modeled background estimate, is higher than 6 at the position of the pulsar PSR J0205+6449. The excess of the VHE skymap was fit with a Gaussian function. The best-fit position is RA(J2000)~=~2~h 05~m 31(09)$_\text{stat}$(11)$_\text{sys}$~s ; DEC~(J2000)~=~$64^\circ$ 51$^{\prime}$(1)$_\text{stat}$(1)$_\text{sys}$. This position is statistically deviant by $2\sigma$ from the position of the pulsar, but is compatible with it if systematic errors are taken into account. In the bottom left of the image we show the point spread function (PSF) of the smeared map at the corresponding energies, which is the result of the sum in quadrature of the instrumental angular resolution and the applied smearing (4.7$^{\prime}$ radius, at the analysis energy threshold). The extension of the VHE source is compatible with the instrument PSF. The VLA contours are coincident with the detected $\gamma$-ray excess.

Figure \ref{SED} shows the energy spectrum for the MAGIC data, together with published predictions for the gamma-ray emission from several authors, and two spectra obtained with three years of {\it Fermi}-LAT data, which were retrieved from the {\it Fermi}-LAT second pulsar-catalog \citep[2PC,][]{SecondPulsarCatalog} and the {\it Fermi} high-energy LAT catalog \citep[1FHL,][]{FermiAbove10}. The 1FHL catalog used events from the {\it Pass 7 Clean class}, which provides a substantial reduction of residual cosmic-ray background above 10 GeV, at the expense of a slightly smaller collection area, compared with the  {\it Pass 7 Source class} that was adopted for 2PC \citep{Clean_class}. The two $\gamma$-ray spectra from 3C58 reported in the 2PC and 1FHL catalogs agree within statistical uncertainties. The differential energy spectrum of the source is well fit by a single power-law function d$\phi$/d$E$=$f_0$($E/$1 TeV)$^{-\Gamma}$ with  $f_{0} = (2.0 \pm 0.4_{\text{stat}} \pm 0.6_{\text{sys}})\ttt{-13} \text{cm}^{-2} \text{s}^{-1} \text{TeV}^{-1}$, $ \Gamma = 2.4 \pm 0.2_{\text{stat}}\pm 0.2_{\text{sys}}$ and $\chi^2$=0.04/2. The systematic errors were estimated from the MAGIC performance paper \citep{Performance2012} including the upgraded telescope performances. The integral flux above 1 TeV is $F_{E>1\text{ TeV}}=1.4 \ttt{-13} \text{cm}^{-2} \text{s}^{-1}$. Taking into account a distance of 2 kpc, the luminosity of the source above 1 TeV is  $L_{\gamma, E>1 \text{ TeV}}=(3.0\pm1.1)\times$10$^{32}$$d^2_{2}$ erg s$^{-1}$, where $d_{2}$ is the distance normalized to 2 kpc.

  \begin{figure}[!t]
 \centering
 \includegraphics[width=0.5\textwidth]{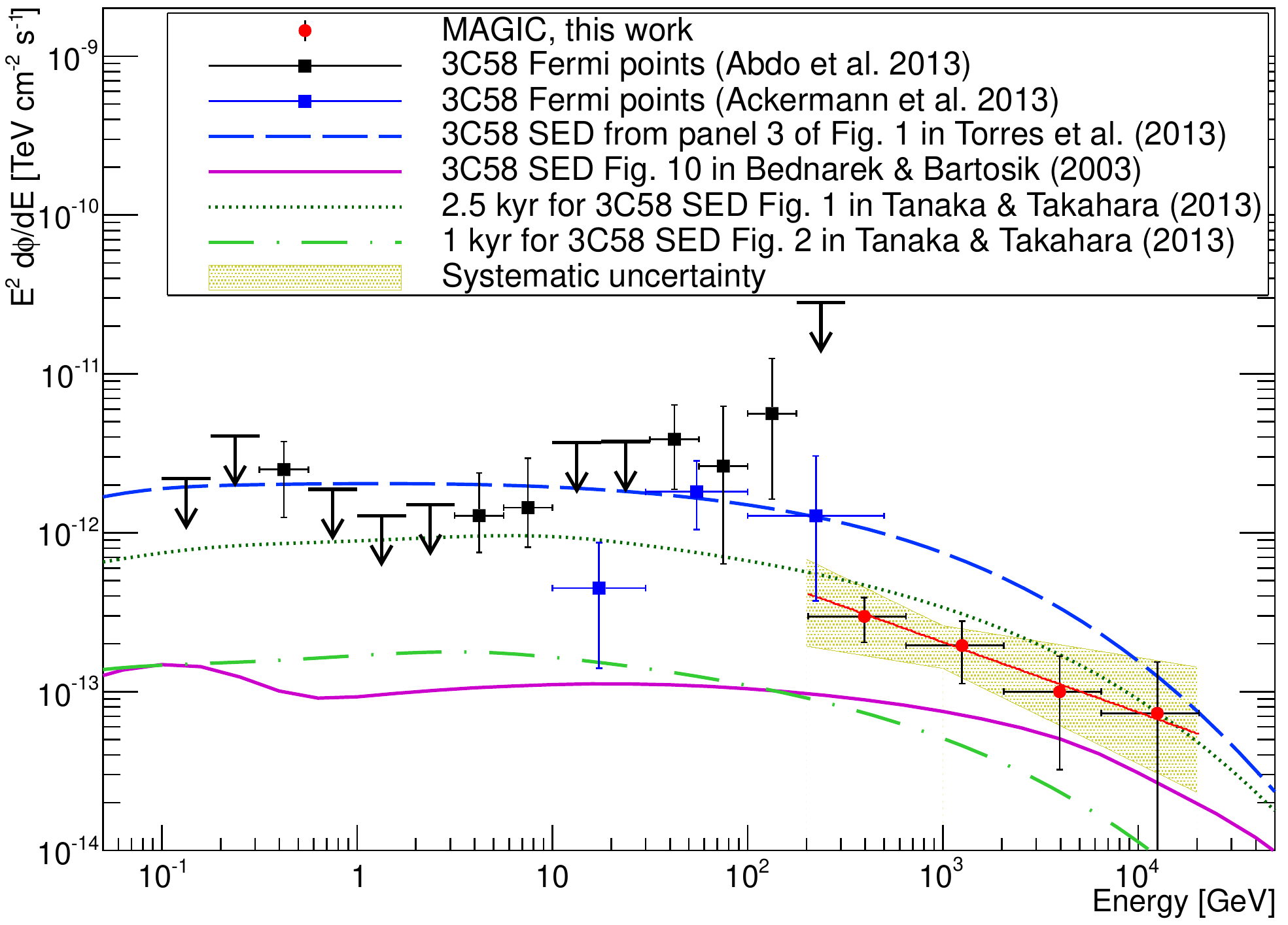}
\caption{3C 58 spectral energy distribution in the range between 0.1 GeV and 20 TeV. Red circles are the VHE points reported in this work. The best-fit function is drawn in red and the systematic uncertainty is represented by the yellow shaded area. Black squares and black arrows are taken from the {\it Fermi}-LAT second pulsar-catalog results \citep{SecondPulsarCatalog}. Blue squares are taken from the {\it Fermi} high-energy LAT catalog \citep{FermiAbove10}. The magenta line is the SED prediction for 3C 58 taken from Figure 10 of \cite{Wlodek2003}. The clear green dashed-dotted line is the SED predicted by \cite{Tanaka13}, assuming an age of 1 kyr, and the dark green dotted line is the prediction from the same paper, assuming an age of 2.5 kyr. The blue dashed line represents the SED predicted by \cite{3C58_Diego} assuming that the Galactic FIR background is high enough to reach a flux detectable by the MAGIC sensitivity in 50h.}
  \label{SED}
 \end{figure}


\section{Discussion}

Several models have been proposed that predict the VHE $\gamma$-ray emission of PWN 3C 58. \cite{Bucciantini} presented a one zone model of the spectral evolution of PWNe and applied it to 3C 58 using a distance of 3.2 kpc. The VHE emission from this model consists of inverse Compton (IC) scattering of CMB photons and optical-to-IR photons, and also of pion decay. The flux of $\gamma$ rays above 400 GeV predicted by this model is about an order of magnitude lower than the observation. 

\cite{Wlodek2003} proposed a time-dependent model in which positrons gain energy in the process of resonant scattering by heavy nuclei. The VHE emission is produced by IC scattering of leptons off CMB, IR, and synchrotron photons and by the decay of pions due to the interaction of nuclei with the matter of the nebula. The age of 3C 58 is assumed to be 5 kyr, using a distance of 3.2 kpc and an expansion velocity of 1000 km s$^{-1}$. According to this model, the predicted integral flux above 400 GeV is $\sim$10$^{-13}$ cm$^{-2}$s$^{-1}$, while the integral flux above 420 GeV measured here is 5$\times$10$^{-13}$cm$^{-2}$s$^{-1}$. Calculations by \cite{Wlodek2005}, using the same model with an initial expansion velocity of 2000 km s$^{-1}$ and considering IC scattering only from the CMB, are consistent with the observed spectrum. However, the magnetic field derived in this case is $B\sim$14$\mu$G and it underestimates the radio emission of the nebula, although a more complex spectral shape might account for the radio nebula emission.

 \cite{Tanaka10} developed a time-dependent model of the spectral evolution of PWNe including synchrotron emission, synchrotron self-Compton, and IC. They evolved the electron energy distribution using an advective differential equation. To calculate the observability of 3C 58 at TeV energies they assumed a distance of 2 kpc and two different ages: 2.5 kyr and 1 kyr \citep{Tanaka13}. For the 2.5 kyr age, they obtained a magnetic field B$\geq17 \mu$G, while for an age of 1 kyr, the magnetic field obtained is B=40 $\mu$G. The emission predicted by this model is closer to the {\it Fermi} result for an age of 2.5 kyr.

\cite{Jonatan} presented a different time-dependent leptonic diffusion-loss equation model without approximations, including synchrotron emission, synchrotron self-Compton, IC, and bremsstrahlung. They assumed a distance of 3.2 kpc and an age of 2.5 kyr to calculate the observability of 3C 58 at high energies \citep{3C58_Diego}. The predicted emission, without considering any additional photon source other than the CMB, is more than an order of magnitude lower than the flux reported here. It predicts VHE emission detectable by MAGIC in 50 hours for an FIR-dominated photon background with an energy density of 5 eV/cm$^3$. This would be more than one order of magnitude higher than the local IR density in the Galactic background radiation model used in GALPROP  \citep[$\sim$0.2 eV cm$^{-3}$;][]{GALPROP}. The magnetic field derived from this model is 35 $\mu$G. To reproduce the observations, a large FIR background or a revised distance to the PWN of 2 kpc are required. In the first case, a nearby star or the SNR itself might provide the necessary FIR targets, although no detection of an enhancement has been found in the direction of the PWN. As we mentioned in Sec. \ref{intro}, a distance of 2 kpc has recently been proposed by \cite{Kothes2013} based on the recent H I measurements of the Canadian Galactic Plane Survey. At this distance, a lower photon density is required to fit the VHE data (Torres. 2014, priv. comm.).

We have shown different time-dependent models in this section that predict the VHE emission of 3C 58. The SEDs predicted by them are shown in Figure \ref{SED}. They use different assumptions for the evolution of the PWN and its emission. \cite{Bucciantini} divided the evolution of the SNR into phases and modeled the PWN evolution inside it. In \cite{Wlodek2003} model, nuclei play an important role in accelerating particles inside the PWN. \cite{3C58_Diego} and \cite{Tanaka10} modeled the evolution of the particle distribution by solving the diffusion-loss equation.  \cite{3C58_Diego} fully solved the diffusion-loss equation, while \cite{Tanaka10} neglected an escape term in the equation as an approximation. Another difference between these latter two models is that \cite{Tanaka10} took synchrotron emission, synchrotron self-Compton and IC into account, while \cite{3C58_Diego} also consider the bremsstrahlung. The models that fit the $\gamma$-ray data derived a low magnetic field, far from equipartition, very low for a young PWN, but comparable with the value derived by \cite{IR_Mag_field} using other data.

We also evaluated the possibility that the VHE $\gamma$ rays might be produced by hadronic emission in the SNR that is associated with 3C 58. For the calculations, we assumed a distance of 2 kpc, a density of the medium of 0.38 $d^{-1/2}_{2}$ cm$^{-3}$ \citep{Slane2004}, and an initial energy of the supernova explosion of 10$^{51}$ erg. We calculated the expected flux above 1 TeV following the approach reported in \cite{Fluxes_SNRs}. We find that the cosmic-ray nuclei have to be accelerated with an efficiency above 100\% to account for our observed flux, thus we conclude that cosmic rays from the SNR are unlikely to be the origin of the TeV emission. Note, however, that dense clumps embedded within the remnant may change these conclusions \citep{Clumps}.


\section{Conclusions}

We have for the first time detected VHE $\gamma$ rays up to TeV energies from the PWN 3C 58. The measured luminosity and flux make 3C 58 the least-luminous VHE $\gamma$-ray PWN known and the object with the lowest flux at VHE to date. Only a closer distance of 2 kpc or a high local FIR photon density can qualitatively reproduce the multiwavelength data of this object in the published models. Since the high FIR density is unexpected, the closer distance with FIR photon density comparable with the averaged value in the Galaxy is favored. The models that fit the $\gamma$-ray data derived magnetic fields which are very far from equipartition. Following the assumptions in \cite{Fluxes_SNRs}, it is highly unlikely that the measured flux comes from hadronic emission of the SNR. The VHE spectral results we presented will help to explain the broadband emission of this source within the available theoretical scenarios.


\begin{acknowledgements}

We would like to thank
the Instituto de Astrof\'{\i}sica de Canarias
for the excellent working conditions
at the Observatorio del Roque de los Muchachos in La Palma.
The support of the German BMBF and MPG,
the Italian INFN, 
the Swiss National Fund SNF,
and the Spanish MINECO
is gratefully acknowledged.
This work was also supported
by the CPAN CSD2007-00042 and MultiDark CSD2009-00064 projects of the Spanish Consolider-Ingenio 2010 programme,
by grant 127740 of the Academy of Finland,
by the DFG Cluster of Excellence ``Origin and Structure of the Universe'',
by the Croatian Science Foundation (HrZZ) Project 09/176,
by the DFG Collaborative Research Centers SFB823/C4 and SFB876/C3,
and by the Polish MNiSzW grant 745/N-HESS-MAGIC/2010/0. 
We would like to thank S. J. Tanaka for providing us useful information about his model.

\end{acknowledgements}

 
\bibliographystyle{aa} 
\bibliography{./aa.bib} 



\end{document}